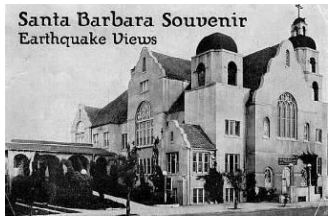

# Effects of Surface Geology on Seismic Motion

August 23–26, 2011 · University of California Santa Barbara

# NUMERICAL MODELLING OF SHAKING EFFECTS DUE TO STRONG MOTIONS ON THE TIBER ALLUVIAL DEPOSITS IN ROME (ITALY)

**Fabian BONILLA**
IFSTTAR
Paris East University
France

**Francesca BOZZANO**
CERI- Dip. Scienze della Terra
University of Rome "Sapienza"
France

**Celine GÉLIS**
IRSN
Paris
France

**Anna Chiara GIACOMI**
Dip. Scienze della Terra
University of Rome "Sapienza"
Italy

**Luca LENTI**
IFSTTAR
Paris East University
France

**Salvatore MARTINO**
CERI- Dip. Scienze della Terra
University of Rome "Sapienza"
Italy

**Maria Paola SANTISI D'AVILA**
Laboratorie de Mécanique des
Solides Ecole Polytechnique
Paris, France

**Jean-Françoise SEMBLAT**
IFSTTAR
Paris East University
France

ABSTRACT

A multidisciplinary approach is proposed for evaluating the effects of shaking due to strong motions on the Tiber river alluvial deposits in Rome's historical centre. At this aim, a detailed 3D geological model of the Tiber river alluvial deposit has been constructed and a numerical analysis of site response was performed along two geological sections across the historical centre of Rome. The numerical models were performed in both 1D and 2D configurations assuming linear and nonlinear conditions, by applying a three component seismic input. The results show that the maximum shear strains are strongly conditioned by the layer geometries (i.e. 2D or 1D conditions) and by the soil heterogeneity. Moreover, the reliability of the maximum strains obtained by numerical modeling is discussed comparing these values respect to both the volumetric and the degradation dynamic thresholds of the considered soils.

INTRODUCTION

The geological setting of the historical centre of Rome is characterised by alluvial deposits, related to the main Tiber river valley and subsequently hidden by a man-made fill, resulting from many centuries of urban settlement. The city of Rome is located at a distance of some tens of kilometres from the central Apennines seismogenic zone, where earthquakes of tectonic origin and of up to 7.0 magnitude can be expected. The most recent of these major earthquakes took place the on April $6^{th}$, 2009 (Mw 6.3) close to L'Aquila city, about 100 km NE of Rome (Blumetti et al., 2009). The mainshock of L'Aquila earthquake was felt at Rome up to V MCS intensity; nevertheless, damages to buildings, widespread located, have been denounced by inhabitants to the local authorities during the weeks that followed the mainshock.

More in particular, the seismogenetic sources of this regional seismicity are located as close as 60-100 km from Rome; they have focal depth of about 10-15 km and can produce strong earthquakes with a magnitude up to 7 (Rovelli et al., 1994). Smaller earthquakes, with a focal depth of ≤ 6 km and maximum magnitude of 5, derive from the Colli Albani hills volcanic source (Amato et al., 1994). Moreover, a local seismicity in the urban area can produce earthquakes with magnitude not exceeding 4 (Tertulliani et al., 1996). These smaller events are expected to produce a maximum $I_{felt}$= VI-VII in Rome.

In terms of peak ground acceleration (PGA), the expected values for a magnitude 7 earthquake, at a distance of 100 km, range from roughly 0.03g to roughly 0.07g. These values are in agreement with the 10% probability of exceedance in 50 years for the PGA, as shown by Romeo at al. (2000).

A study by Ambrosini et al. (1986) reported that the 1915 earthquake caused the most severe damage to buildings located on the Holocene alluvial fill of the Tiber valley, thereby stressing the role of local geology in amplifying ground shaking. Similarly, Boschi et al., (1995) and Funiciello et al. (1995) showed a close relationship between selective damage to monuments, such as the Antonina Column and the Colosseum, and presence of soft alluvial sediments underneath.

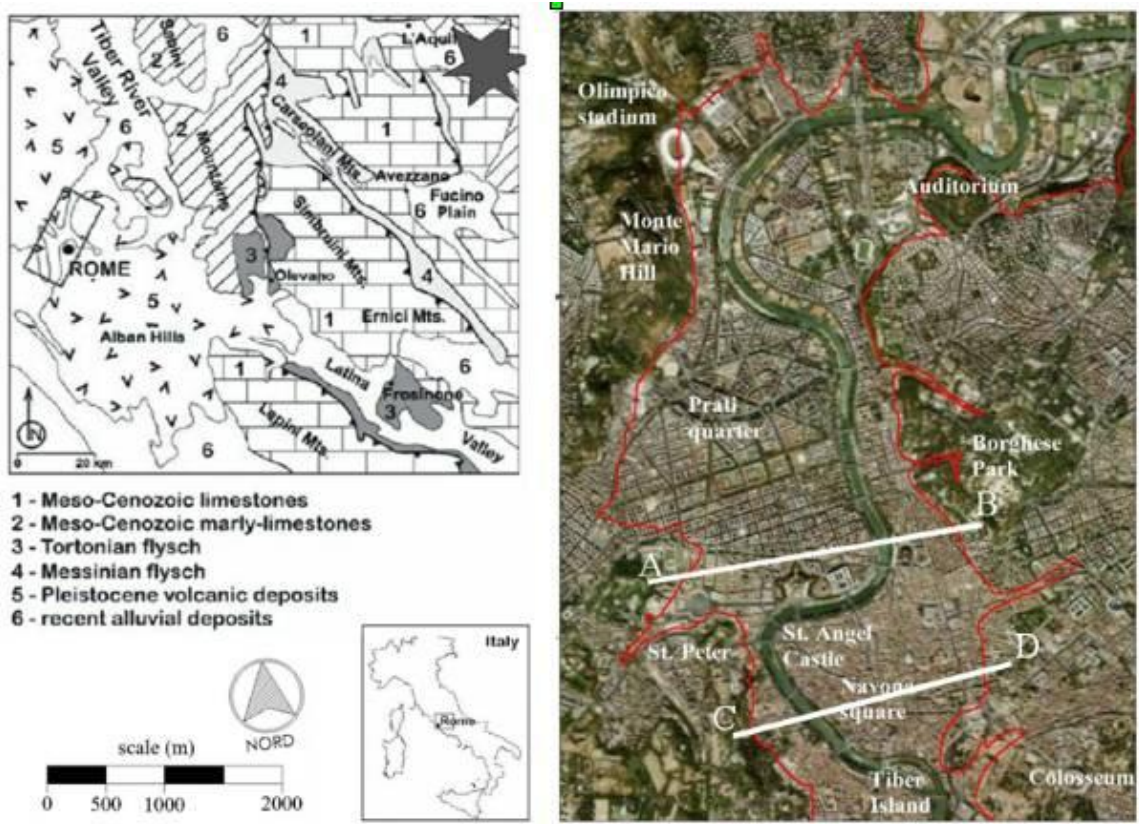

*Fig. 1. Location of the city of Rome in the central Apennines area; the epicenter of the last 6$^{th}$ April, 2009 earthquake is also shown (left). Lateral boundary of the Tiber river alluvial deposits in correspondence with the historical center of Rome (red line) and traces of the geological sections studied here (right).*

In the last decades numerous studies were conducted to quantify the expected site effects within the city of Rome due to the basin-like geometry of the Tiber river valley as well as to its alluvial fill. In this regard, Rovelli et al. (1994; 1995) performed 2D simulations via the finite-difference technique and a hybrid technique based on summation and finite difference was proposed by Fah at al. (1993). Olsen et al. (2006) proposed a new 3D subsurface model of Rome, but with a homogeneous fill. They pointed out that the alluvial fill within the main Tiber River valley has one principal mode of vibration at about 1 Hz; nevertheless, nor the heterogeneity of these alluvia neither the nonlinear effects due to the shaking were taken into account.

Static and dynamic geomechanical properties of the Holocene alluvial fill within the Tiber River valley were provided by Bozzano et al. (2000; 2008). The Authors also demonstrated that the silty-clay deposits, which constitute the most part of the Tiber alluvial body, play a key role in characterizing the deformation profile along the soil column.

Previous numerical simulations, performed in both 1D and 2D dimensions and referred to large urban areas (Bonilla et al., 2010; Bonilla et al., 2006; Bouden-Romdhane et al., 2003), were devoted to analyse possible local effects due to the modification of the input seismic wavefield in the superficial layers, both in linear and nonlinear conditions. These numerical models demonstrated that the basin response depends on many elements such as site geometry, impedance contrast, material elastic and dynamic properties, as well as on the stress field induced by the seismic motion that may lead to relevant nonlinear effects. In particular, the results proved that the stronger is the impedance contrast between the sediments and the surrounding bedrock, the stronger the amplification is, regardless of the basin geometry. As a consequence, the amplifications resulting from 2D conditions are significantly higher than the ones obtained under 1D conditions. In particular, 1D and 2D numerical modeling of local seismic response highlight relevant effects due to the heterogeneities of the alluvia, both in terms of amplification functions as well as stress-strain resulting distribution.

Since the shape ratio of the Tiber river valley close to the Rome's historical centre is higher than the one requested for a proper 2D resonance (Bard and Bouchon, 1985; Rovelli et al., 1994), the alluvial fill and its heterogeneity mainly controls the local seismic response (Bonilla et al., 2010). In the frame of these researches, the present study proposes a multidisciplinary approach aiming at evaluating the effects due to nonlinearity in the alluvial fill of the Tiber river in Rome's historical centre.

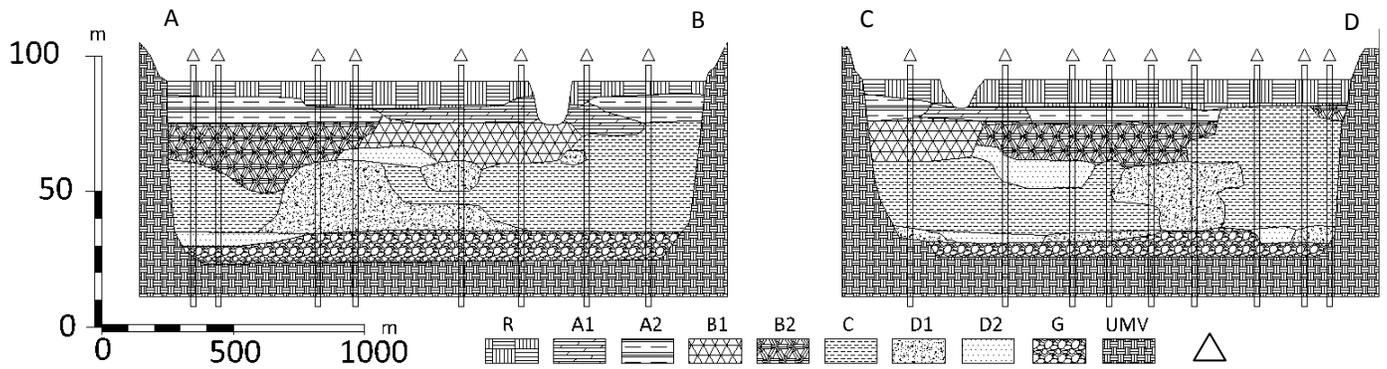

*Fig. 2. Engineering-geology model along sections AB and CD of Fig.1: triangles indicate the considered boreholes.*

A detailed 3D engineering-geological model was considered for deriving geological sections to be modelled by the use of 2D and 1D nonlinear numerical codes, in order to analyze both the seismic amplification and the maximum shear strains, whose distribution was compared with the geological setting of the alluvial deposits and whose values were compared with the dynamic thresholds of the alluvial deposits.

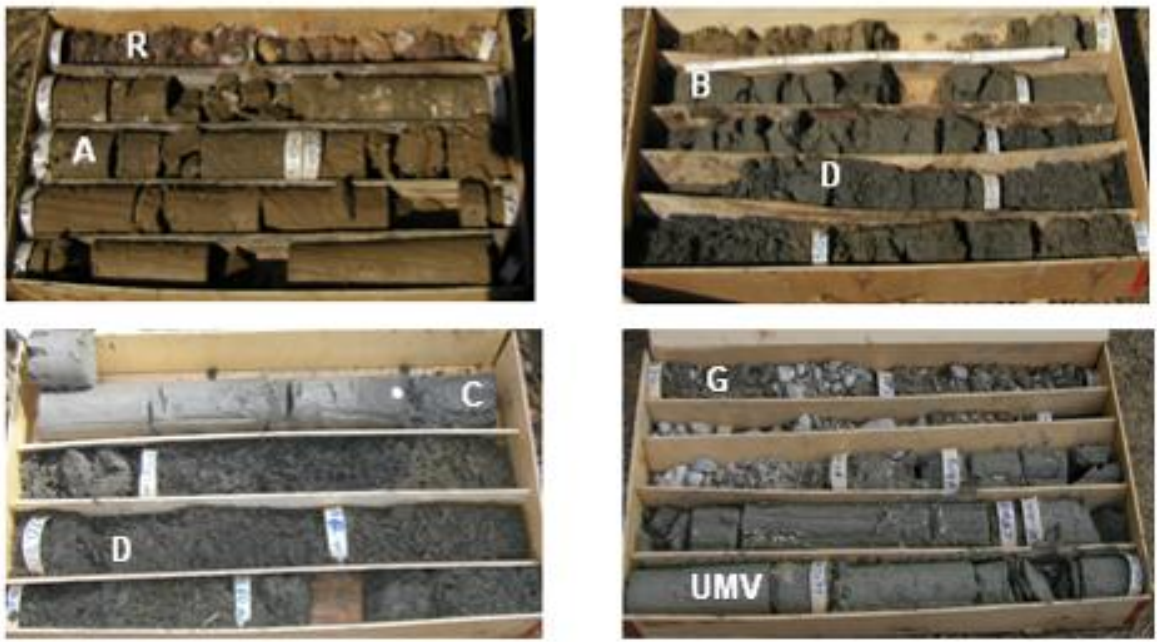

*Fig. 3. Alluvial deposits of the Tiber river in the Rome's historical centre from borehole sampling.*

GEOLOGICAL SETTING OF THE CITY OF ROME

The area of Rome was characterised by marine sedimentary conditions from Pliocene through early Pleistocene times (4.5-1.0 Myr). This Plio-Pleistocene succession consists of alternating, decimetre-thick levels of clay and sand, with an overconsolidation ratio (OCR) >5 and low compressibility (Bozzano et al., 1997). The Pliocene high consistency clays, ascribed to the Monte Vaticano Unit (MVU), represent the seismic bedrock of the area of Rome. During middle-late Pleistocene and Holocene times, sedimentary processes were confined to fluvial channels and coastal plains and strongly controlled by glacio-eustatic sea-level changes (Karner and Renne, 1998; Karner and Marra, 1998, Marra et al., 1998). Moreover, the present-day hydrographic network of the Tiber valley and its tributaries originated from the Würm glacial period (18 kyr). The alluvial deposits which fill the Holocene valleys are generally characterised by a succession, with a relatively thin level of gravel at the base grading into a thick pack of sand and clay (Bozzano et al., 2000); this fine-grained portion of the deposits is represented by normally to weakly overconsolidated clayey and sandy silt, saturated in water, with low rigidity moduli (Bozzano et al., 2000).

A 3D engineering-geological model of the alluvial deposits in correspondence with the historical center of Rome has been constructed, based on data stored from 78 continously-cored boreholes; these boreholes, which range in depth from 30 to 67 m b.g.l., penetrate through the sediments filling the Tiber river valley and, in many cases, terminate in the UMV clays (Fig.3).

The 3D stratigraphic reconstruction of the alluvial deposits shows that the Tiber valley in Rome's historical centre has a depth varying from 60 to 70 m moving from north to south. The valley, which is enclosed by steep slopes, has a flat floor slightly dipping southwards (Fig. 2).

According to the 3D geological model, basal gravels in a sandy-silty matrix (level G) cover the UMV with a thickness ranging between 5 and 10 m. The main alluvial deposits can be ascribed to levels C and D. In particular, the C level is constituted of silty clays with a variable organic content and reaches a maximum thickness of about 50 m (see section AB of Fig.2), while the level D is composed by silty-sands passing to clayey-silts and it can be divided in two sub-levels. Sub-level D1 is characterized by a larger grain size and sub-level D2 is characterized by finer grain content. Level B is composed by sands (B1), which laterally pass to silty sands and clays (B2). The historical alluvia of the Tiber (level A) close the sedimentary succession, ascribed to the natural deposits of the Tiber river path; this level is composed of silty-sands locally passing to clayey silts. Also in this case two sub-levels can be distinguished, based on their grain sizing: a sandy sub-level A1 and a clayey sub-level A2. Finally, human fills (level R), overly the historical alluvia of the Tiber (level A) and are characterized by abundant, variously sized brick fragments and blocks of tuff embedded in a brown-green silty-sandy matrix. This horizon also contains ceramic and mortar fragments. The average thickness the R level close to the historical centre of Rome is of about 7 m. Two geological sections (AB and CD of Fig. 2) were derived from the 3D engineering-geological model; these sections are almost W-E oriented, about 2 km long, and across the Tiber river valley in correspondence with the historical centre of Rome.

NUMERICAL MODEL

Engineering-geological model

An engineering-geological model was reconstructed sections AB and CD of Fig.2. At this aim, the values of the physical properties and of the index parameters (Table 1) are attributed according to Bozzano et al. (2008). On the other hand, the dynamic mechanical properties of the Tiber alluvial deposits were derived by resonant column (RC) and cyclic torsional shear tests (Bozzano et al., 2008). The derived linearity threshold for the shear strains corresponds to about 0.005% for the UMV and to 0.01% - 0.02% for the alluvial deposits, while the volumetric threshold for the alluvia ranges from 0.02% to 0.04%.

Site and laboratory tests on the Tiber alluvial deposits (Fig.4, Tab.1), proved that the highest impedance contrast occurs between levels A, B, C, D (average Vs of about 300 m/s) and the basal level G (Vs = 713 m/s). According to Bozzano et al. (2008), the G level can be regarded as the seismic bedrock for the historical centre of Rome. On the contrary, the UMV (Vs = 545 m/s), which can be regarded as the geological substratum of the Tiber alluvia in Rome, causes an inversion of the Vs value along the vertical soil profile, corresponding to a low velocity difference with respect to the G level (i.e. Vs difference lower than 200 m/s). Moreover, the relatively low values of Vs (<600 m/s), measured within the first 10 m of UMV, can be referred to depending a "softened stratum" which is related to the stress release caused by the Holocene fluvial erosion (Bozzano et al., 2006).

In order to take into account this specific feature of the Rome's geological substratum, a linear increasing gradient of Vs (i.e. from 540 m/s - laboratory test datum - up to 1000 m/s with thickness of 20 m) was assumed, in the numerical models, within the UMV. Moreover, an hysteretic constitutive law, according to the resonant column tests, was attributed to the UMV portion characterized by Vs< 800 m/s; whereas a visco-elastic constitutive law was attributed to the remnant portion of the UMV deposit.

Seismic input

A three-components seismic input was derived (in co-operation with Dott. Guido Martini, ENEA-Frascati, Italy), according to the maximum PGA expected in the historical centre of Rome. In this regard, an historical analysis of seismic response for the city of Rome was performed in order to obtain a couple of magnitude/distance values for each seismic event. Starting from these parameters, the European Strong-motion Database (ESD) was investigated, and a second selection was performed to select a natural three components time history, which is "compatible" with the local site spectrum, expected by UHS INGV - Cluster 6 (Convention INGV-DPC 2004-2006).

The chosen time history was normalizes to the PGA and then scaled to $Ag_0 = 0.1258g$ which can be computed at 475 years of return time for the historical centre of Rome.

Material rheology

For the here performed 1D numerical modeling, the nonlinear finit element code SWAP_3C (Santisi d'Avila et al., 2010) was used and the three components of the seismic motion are propagated into a multi-layered column of nonlinear soil from the top of the underlying elastic bedrock, by assuming a finite element scheme; properties were attributed according to the mechanical and dynamic parameters of Fig.4 and Tab.1.

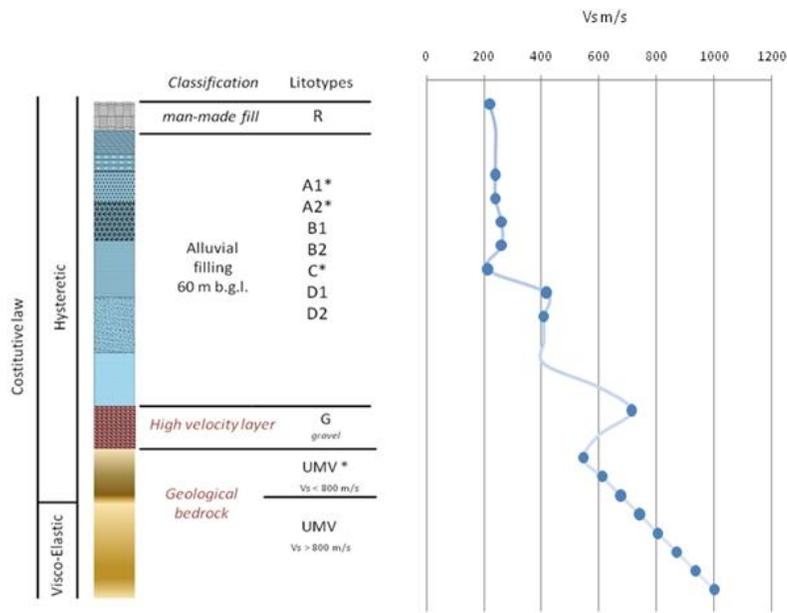

Fig. 4. Schematic soil column of the Tiber alluvia and of the Pliocene substratum, stars indicates the lab-tested soils.

Table 1. Dynamic parameters for levels of Fig.3

| Lithotype | ρ (kg/m³) | $G_0$ (Mpa) | Vp (m/s) | Vs (m/s) |
|---|---|---|---|---|
| R | 1.834 | 88.8 | 490 | 220 |
| A1 | 1.875 | 107.1 | 478 | 239 |
| A2 | 1.875 | 107.1 | 523 | 239 |
| B1 | 1.865 | 126.1 | 1480 | 260 |
| B2 | 1.967 | 132.9 | 1480 | 260 |
| C | 1.865 | 84.23 | 1235 | 212.5 |
| D1 | 1.957 | 340.3 | 1760 | 417 |
| D2 | 2.057 | 340.75 | 1760 | 407 |
| G | 2.14 | 1088.2 | 2560 | 713 |
| UMV | 2.078 | 617.3 | 2125.5 | 545 |
| UMV | 2.078 | 773.4 | 2379 | 610 |
| UMV | 2.078 | 947 | 2632.5 | 675 |
| UMV | 2.078 | 1138.1 | 2886 | 740 |
| UMV | 2.078 | 1346.9 | 3139.5 | 805 |
| UMV | 2.078 | 1573.2 | 3393 | 870 |
| UMV | 2.078 | 1817 | 3646.5 | 935 |
| UMV | 2.078 | 2078.4 | 3900 | 1000 |

Soil is assumed to be a continuous and uniform medium of infinite horizontal extent, while the stratification is discretized into a system of horizontal layers, parallel to the *xy* plane by using quadratic line elements with three nodes. Shear and pressure waves propagate vertically in the *z* – direction; these hypotheses yield no strain variation in *x* - and *y* -directions.

According to a finite element modeling of a multilayer soil system assumed with an horizontal setting, the weak form of equilibrium equations, including compatibility conditions, three-dimensional nonlinear constitutive relation and the imposed boundary condition (Cook et al., 2002), is expressed in matrix form as:

$$MD' + CD'' + F_{INT} = F$$

where M is the mass matrix, D' and D'' are the first and second temporal derivatives of the displacement vector D, respectively, $F_{INT}$ is the vector of internal forces and F is the load vector. C is a matrix that derives from the fixed boundary condition.

The system of horizontal soil layers is bounded at the top by the free surface and at the bottom by the semi-infinite elastic medium which represents the seismic bedrock. The stresses normal to the free surface are imposed null and at the interface soil-bedrock the following condition, implemented by Joyner and Chen (1975) in a finite difference formulation; this model takes into account the nonlinear hysteretic behavior of soils, using an elasto-plastic approach with hardening based on the definition of a series of nested yield surfaces (Iwan, 1967). The main feature of this rheological model is its flexibility for incorporating laboratory results on the dynamic behavior of soils. The 2D numerical models were performed by a 2D P-SV finite difference code using a stencil proposed by Saenger et al. (2000). This stencil allows computing all components of the stress-strain tensor in one point of the numerical mesh, which simplifies the numerical evolution of nonlinear soil rheologies. Wave propagation in heterogeneous linear and nonlinear media is efficiently modelled. Furthermore, the free surface is easily introduced by zeroing Lamé parameters above the free surface and surface waves can be modelled more accurately (Gélis et al., 2005) than with traditional staggered-grid methods (Virieux, 1986).

The model total size is 90 m depth and almost 2 km length for the 2 profiles; nevertheless, the domain corresponding to the basins of AB and CD sections were laterally extended. Furthermore, absorbing boundary conditions are guaranteed at the bottom and the sides of the model.

The minimum and maximum shear wave velocities in linear conditions are 210 and 1000 m/s for the alluvial deposits and the bedrock respectively (Bozzano et al., 2008). The P wave velocities were deduced from the measured geomechanical properties (Fig.4 and Table 1). The numerical spatial and time steps were 0.5 m and 5.0E-5 s which permit to have reliable results in linear and nonlinear simulations up to 10 Hz. The minimum quality factors for S waves ($Q_S$) are computed, when possible, from the experimental values of the damping ratio, measured by resonant column tests at small deformations. The values for $Q_P$ were assumed equal to $2Q_S$. In this study, we introduce attenuation for all linear simulations by using the method proposed by Liu and Archuleta (2006). The strain-stress relation, governing the nonlinear behaviour modelling, is based on the multishear mechanism model by Towhata and Ishihara (1985).

The multishear mechanism model is a plane strain formulation to simulate pore pressure generation in sands under cyclic loading and undrained conditions. After the work by Iai at al. (1990a and 1990b), the model was modified to account for the cyclic mobility and dilatancy of sands. However, in its basic form, this formulation models soil nonlinearity without taking into account any contribution from co-seismic water pore pressures.
For more details on the nonlinear stress-strain rheology, the reader may see the papers by Iai et al. (1990a and 1990b).
The wave motion is computed based on a combination of viscoelastic and nonlinear rheologies. First, the stress tensor components are calculated with the viscoelastic rheology of Liu and Archuleta (2006) and equivalent strain tensor components are deduced. We then calculate the stress tensor components using the nonlinear rheology of Iai et al (1990a and 1990b) to get the stress tensor components. This two-step computation allows taking into account anelastic losses (that are dominant for weak motion) and nonlinear soil behaviour (that controls strong motion). This requires subtracting to the damping curves the part already taken into account through anelastic losses ($Q_S$).
The input was applied as a vertical upward SV planar wave, located within the 2D stratigraphy, at a depth of 90 m from the surface, and at the base of the soil columns for the 1D simulations. The source we used is a synthetic waveform simulating a M7 earthquake with an epicentral distance in the range 80-100 km referred to the Apennines seismogenic zone and with a corresponding PGA of 0.13g referred to the outcropping bedrock (Rovelli et al., 1994).
The reference accelerogram was divided by 2 to take into account the free surface effect. For the 2D simulations, the Fast Fourier Transform (FFT) of the velocity time histories obtained at the model surface along the entire horizontal domain were computed. The reference rock outcropping motion was computed with a 1D model corresponding to the soil column outside the basin (Vs at surface equal to 545 m/s).
The spectral ratios to this reference motion were obtained from the synthetic accelerations all along the modelled section. In addition, maximum stress and strains are also calculated at each point of the numerical domain. Finally, for the 1D simulations, the reference signal is directly the reference outcropping accelerogram, outcropping rock properties correspond to the bedrock ones (Vs=1000 m/s).

RESULTS

The obtained results point out some interesting insights on: 1) the role of heterogeneity in nonlinear modelling by 2D and 1D numerical simulations; 2) the reliability of the results in terms of maximum shear strains obtained by numerical modelling under nonlinear simulations if compared with the dynamic soil thresholds, measured through lab-tests.
About the role of heterogeneity in nonlinear modelling the here reported outputs highlight how the juxtaposition of the plastic layers within the stiffer ones (i.e. "boxed" layers) can significantly increase the values of the modeled maximum shears strains (Fig. 6); on the other hand, numerical 1D modeling generally underestimate the maximum shear strains with respect to the 2D ones (Fig. 6-7).
The results in terms of shear strains obtained by numerical modeling demonstrate that the significant exceedance (up to one order of magnitude) of the volumetric threshold under nonlinear conditions involves the main part of the model domain corresponding to the alluvial deposits. In this regard, it is worth noting that the volumetric threshold exceedance can guarantees a more conservative evaluation of the nonlinear field of reliability of numerical models which do not simulate the generation of exceeding pore pressures; nevertheless, according to Diaz-Rodriguez and Lopez-Molinas (2008) the "degradation threshold" actually represents the reference value of the shear strain level which marks the beginning of the decisive de-structuring phase of a soil specimen during the laboratory dynamic tests. In fact, for normally consolidated clayey soil, the values of the degradation strain thresholds can be fixed in the range 0.5% up to 3-5% (Houston and Herrmann 1980, Lefevbre et al. 1989, Díaz-Rodríguez and Santamarina 2001). On the other hand, the strain values referred to the volumetric threshold, lab-derived for deposits similar to the ones considered here (Sun et al., 1988; Carrubba and Maugeri, 1988; Ishibashi and Zhang 1993; Boulanger et al., 1998, Crespellani et al., 2001), are generally of one order of magnitude lower, i.e. they vary in the range 0.01% up to 0.1%. As a consequence, the Authors' idea is that the nonlinear rheological behavior adopted here for the Tiber river fill deposits can be actually regarded as reliable for providing shear strains without considering exceeding pore pressures up to the dynamic degradation threshold.
In any case, future applications of 1D and 2D nonlinear codes used here should be done taking into account the generation of exceeding pore pressures and, as a consequence, they will able to provide results in terms of computed strains, also above the dynamic degradation threshold. In this regard, some analogical and not conventional experiments, just referred to the Rome's case study, will be performed in the frame of the SERIES project ENINALS (Experimental and Numerical Investigations of Nonlinearity in soils using Advanced Laboratory-Scaled models) at the IFSTTAR's centrifuge device of Nantes, in order to provide specific data on the rheological behaviour at large shear strain levels by monitoring exceeding pore pressure within multilayered soil columns.

CONCLUSIONS

Two geological sections were obtained across the Tiber river fill in the historical centre of Rome, based on a 3D detailed engineering-geological model. 1D and 2D numerical modelling, under linear and nonlinear conditions, were performed along these sections in order to analyze the local seismic response as well as to evaluate the possible role of heterogeneity in the resulting maximum shear strains. The obtained numerical outputs confirm a main resonance frequency of the alluvial fill close to 1Hz but, at the same time, they show significant amplifications in the frequency range 3-8Hz, due to the heterogeneity of the alluvia. This has not been shown before by previous studies.

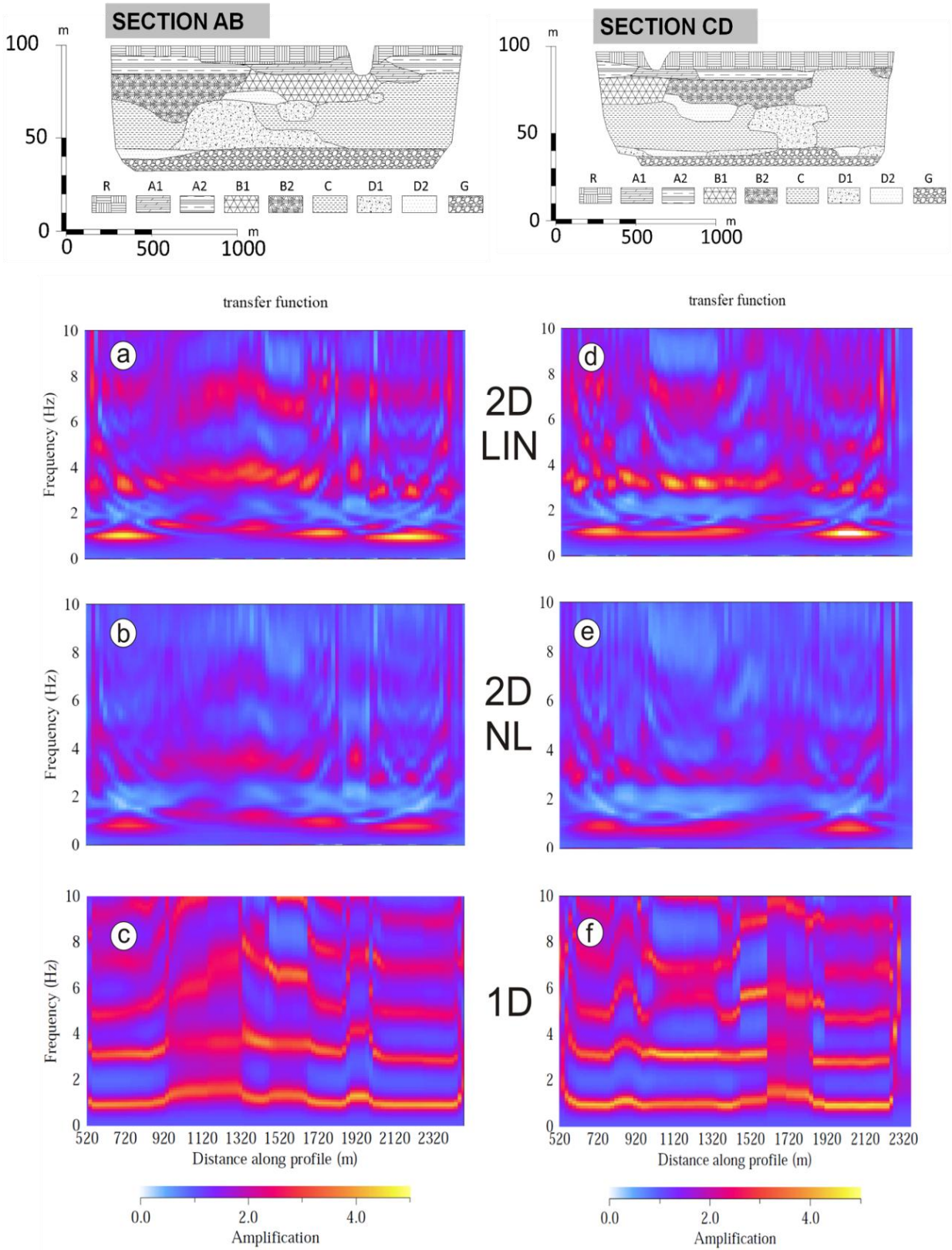

*Fig. 5. Comparison between amplification functions resulting by numerical modeling for the sections AB (on the left) and CD (on the right) of Figs. 1 and 2 under linear-2D (a-d), nonlinear-2D (b-e) and linear-1D (c-f) conditions; the reference engineering-geology sections are also reported.*

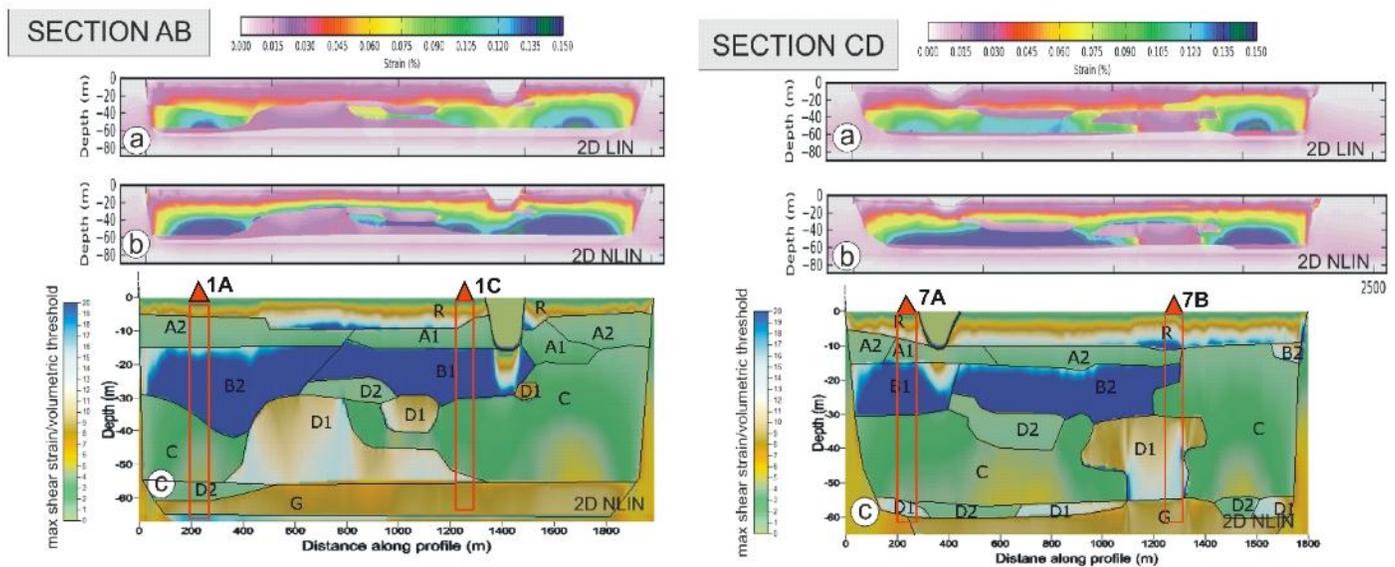

*Fig. 6. Maximum shear strains resulting from the numerical modeling along sections AB and CD of in Figs. 1 and 2 under linear-2D (a), nonlinear-2D (b) conditions. The ratio between the maximum shear strain and the dynamic volumetric threshold (c) is also contoured. The red boxes indicate the location of the 4 columns of Fig.7, which were considered for the 1D-nonlinear simulation performed by code SWAP_3C.*

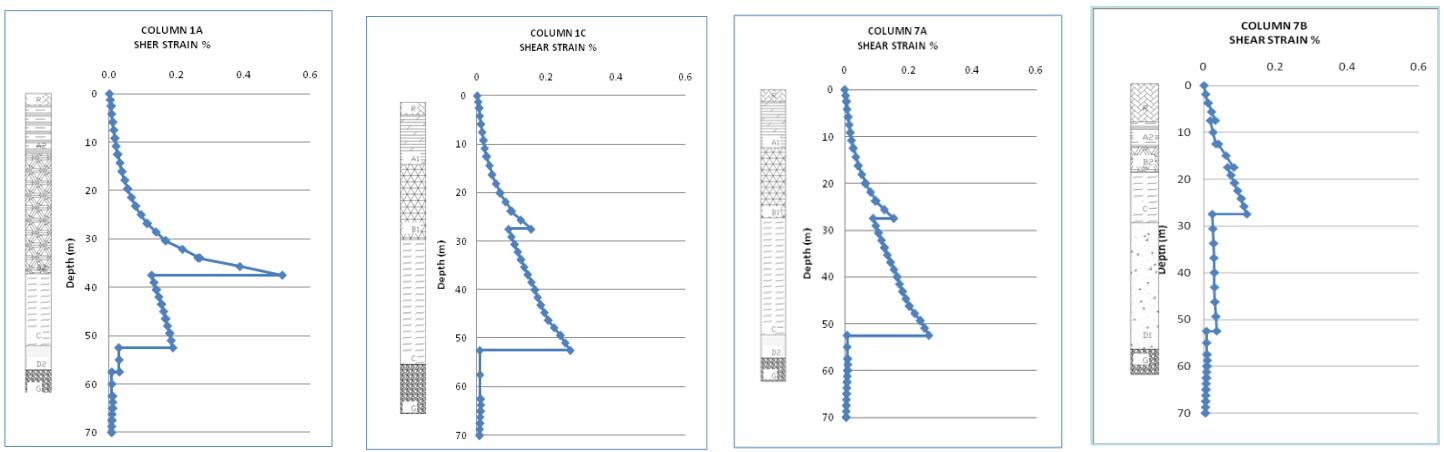

*Fig. 7. Maximum shear strains resulting, along the soil columns of Fig.6, from the nonlinear 1D numerical modeling performed by code SWAP_3C.*

Moreover, the numerical simulations give interesting insights on possible effects due to nonlinearity on the maximum shear strains. The numerical results illustrate that: i) the highest values of the maximum shear strain are reached within plastic layers which are "boxed" within stiffer ones; ii) the dynamic volumetric soil thresholds are exceeded within the main part of the deposits: nevertheless, the reliability of the shear strain values obtained by numerical modelling could be preserved if a degradation threshold is considered as limit shear strain value for the applicability of numerical modelling under nonlinear conditions (i.e., stiffness and damping decay) with no pore pressure generation; iii) 2D geometries are responsible for higher maximum shear strains with respect to the 1D simulation performed by considering a similar nonlinear behaviour.

Finally, because of the very high variability of the soil layering within the Tiber river alluvial fill in Rome, there is need of numerical modelling of other sections. They can be derived from the available 3D engineering-geological model in order to confirm the results obtained here. Furthermore, they could be useful to deduce some correlations between the layering juxtapositions and the expected shear strains, as well as to define indexes for quantifying the fill heterogeneity and its effect on the ground shaking. .

REFERENCES

Amato, A., Chiarabba, C., Cocco, M., di Bona, M. and G. Selvaggi [1994], "The 1989-1990 seismic swarm in the Alban Hills volcanic area, central Italy", Jour. of Volc. and Geoth. Res., 61, 225-237.

Ambrosini, S., S. Castenetto, F., Cevolan, E., Di Loreto, R., Funiciello, L., Liperi, and Molin, D. [1986], "Risposta sismica dell'area urbana di Roma in occasione del terremoto del Fucino del 13-1-1915", Mem. Soc. Geol. It., 35, 445-452.

Bard, P.Y., and M. Bouchon [1985], "The two-dimensional resonance of sediment-filled valleys", Bull. Seism. Soc. Am., 75, 519-541.

Blumetti, A.M., Comerci, V., Di Manna, P., Guerrieri, L. and E. Vittori, [2009], "Geological effects induced by the L'Aquila earthquake (6 April 2009, Ml = 5.8) on the natural environment". ISPRA – Dipartimento Difesa del Suolo - Servizio Geologico d'Italia, preliminary report, 38.

Bonilla, L.F., Liu, P.C. and S. Nielsen [2006], "1D and 2D linear and nonlinear site response in the Grenoble area". *Third International Symposium on the Effects of Surface Geology on Seismic Motion* Grenoble, France, 30 August - 1 September 2006, Paper Number: 082/S02

Bonilla F., Bozzano F., Gélis C., Giacomi A. C., Lenti L., Martino S., and J-F. Semblat [2010], " Multidisciplinary study of seismic wave amplification in the historical center of Rome*", Proc. 5th International Conference on Recent Advances in Geotechnical Earthquake Engineering and Soil Dynamics*, may 24-29 in 2010, paper n° 3.09b, ISBN 1-887009-15-9, S. Diego (California).

Boschi, E., Caserta, A., Conti, C., Di Bona, M., Funiciello, R., Malagnini, L., Marra, F., Martines, G., Rovelli, A. and S. Salvi [1995], "Resonance of subsurface sediments: an unforeseen complication for designers of roman columns", Bull. Seism. Soc. Am., 85, 320-324.

Bouden-Romdhane, N., Kham, M., Semblat, J.F. and P. Mechler, [2003], "1D and 2D response analysis vs experimental data from Tunis city", Beşinci Ulusal Deprem Mühendisliği Konferansı, 26-30 Mayıs 2003, *İstanbul Fifth National Conference on Earthquake Engineering*, 26-30 May 2003, Istanbul, Turkey, Paper No: AE-051

Bozzano, F., Andreucci, A., Gaeta, M. and Salucci R. [2000], "A geological model of the buried Tiber River valley beneath the historical centre of Rome". Bull. Eng. Geol. Env., 59, 1-21.

Bozzano, F., Caserta, A., Govoni, A., Marra, F. and S. Martino [2008], "Static and dynamic characterization of alluvial deposits in the Tiber River Valley: new data for assessing potential ground motion in the city of Rome". Journal of Geophysical Research, 113, B01303, doi: 10.1029/2006JB004873.

Bozzano, F., Funiciello, R., Gaeta, M., Marra, F., Rosa, C. and G. Valentini [1997], "Recent alluvial deposit in Rome (Italy): morpho-stratigrafic, mineralogical and geomechanical characterisation", *In the Proceedings of the International Symposium of Egineering Geology and Environment*, Publ 1, 1193-1198.

Bozzano, F., Martino, S, and M. Priori [2006], "Natural and man-induced stress evolution on slopes: the Monte Mario hill in Rome", Environmental Geology, 50, 505-524.

Boulanger, R.W., Arulnathan, R., Harder, L.F., Torres, R.A., and M.W. Driller, [1998], "Dynamic properties of Sherman Island Peat",Journal of Geotechnical and Geoenvironmental Engineering, 1998, 12-20.

Carrubba, P. and M. Maugeri [1988], "Determinazione delle proprietà dinamiche di un'argilla mediante prove di colonna risonante",

Rivista Italiana di Geotecnica, 22(2), 101-113.

Cook, R. D., Malkus, D. S., Plesha, M. E. and R. J. Witt [2002], "*Concepts and applications of finite element analysis*", John Wiley & Sons, New York, United States.

Crespellani, T., Madiai, C., Simoni, G. and G. Vanucchi (2001), "Dynamic geotechnical testing and seismic response analyses in two sites of the Commune of Nocera Umbra, Italy", Rivista Italiana di Geotecnica, 35(4), 39-52.

Diaz-Rodriguez J.A., and J.A. Lopez-Molina [2008], "Strain thresholds in soils dynamics", *In the Proceedings of The 14th World Conference on Earthquake Engineering, October 12-17, 2008, Beijing, China*

Diaz-Rodriguez, J.A., and C. Santamarina, [2001], "Mexico City soil behavior at different strains: observation and physical interpretation", ASCE Journal of Geotechnical and Geoenvironmental Engineering 127 (9): 783-789.

Fah, D., Iodice, C., Suhadolc, P., and G. F. Panza [1993], A "New Method for the Realistic Estimation of Seismic Ground Motion in Megacities: the Case of Rome", Earthquake Spectra, 9, 643-668.

Funiciello, R., Lombardi, L., Marra, F. and M. Parotto [1995], "Seismic damage and geological heterogeneity in Rome's Colosseum area: are they related?", Annali di Geofisica, 38(3), 267-277.

Gélis C., Leparoux D., Virieux J., Bitri A., Operto S. and G. Grandjean [2005], "Numerical modelling of surface waves over shallow cavities", Journal of Environmental and Engineering Geophysics, 10, 49-59.

Houston, W.N. and H.G. Herrmann [1980], Undrained cyclic strength of marine soils. ASCE Journal of the Geotechnical Engineering Division, 106 (6): 691-712.

Iai, S., Matsunaga Y., and T. Kameoka [1990a], "Strain Space Plasticity Model for Cyclic Mobility", Report of the Port and Harbour Research Institute, 29, 27-56.

Iai, S., Matsunaga Y. and T. Kameoka [1990b], "Parameter Identification for Cyclic Mobility Model", Report of the Port and Harbour Research Institute, 29, 57-83.

Ishibashi, I. and X. Zhang [1993], Unified dynamic shear moduli and damping ratios of sand and clay, Soils and Foundations, 33(1), 182-191.

Iwan W. D. [1967], "On a class of models for the yielding behaviour of continuous and composite systems", J. of Applied Mechanics, No. 34, pp. 612-617.

Joyner, W. B., and A. T. F. Chen [1975], "Calculation of nonlinear ground response in earthquakes", Bull. Seism. Soc. Am., No. 65(5), pp. 1315-1336.

Karner, D.B. and Marra F. [1998], "Correlation of Fluviodeltaic Aggradational Sections with Glacial Climate History: A Revision of the Classical Pleistocene Stratigraphy of Rome", Geol. Soc. Am. Bull., 110, 748-758.

Karner, D.B. and P.R. Renne [1998]. "40Ar/39Ar Geochronology of Roman Volcanic Province Tephra in the Tiber River Valley: Age Calibration of Middle Pleistocene Sea-Level Changes", Bull. Seis. Soc. Am., 110, 740-747.

Lefebvre, G.S., Leboeuf, D., and B. Demers, [1989], Stability threshold for cyclic loading of saturated clay, Canadian Geotechnical Journal 26 (1): 122-131.

Liu, P., and R. J. Archuleta [2006], Efficiently modeling of Q for 3D numerical simulation of wave propagation, Bull. Seism. Soc. Am. 96, 1352-1358, doi:10.1785/0120050173.

Marra, F., Florindo, F., and D.B. Karner [1998]. "Paleomagnetism and geochronology of early Middle Pleistocene depositional sequences near Rome: comparison with the deep sea 18O climate record", Earth and Planetary Science Letters, 159, 147-164.

Olsen, K.B., Akinci, A., Rovelli, A., Marra, F. and L. Malagnini, [2006]. "3D ground-motion estimation in Rome, Italy". Bull. Seism. Soc. Am., 96(1), 133-146.

Romeo, R., Paciello, A. and D. Rinaldis [2000], "Seismic hazard maps of Italy including site effects". Soil Dyn. and Earth. Eng., 20, 85–92.

Rovelli, A., Caserta, A., Malagnini, L., and F. Marra [1994], "Assessment of potential strong motions in the city of Rome", Annali di Geofisica, 37, 1745-1769.

Rovelli, A., Malagnini, L., Caserta, A., and F. Marra [1995], "Using 1-D and 2-D modelling of ground motion for seismic zonation criteria: results for the city of Rome", Annali di Geofisica, 38(5-6), 591-605.

Saenger E., Gold N., and S. Shapiro [2000], "Modeling the propagation of elastic waves using a modified finite-difference grid",Wave Motion, 31, 77-82.

Santisi d'Avila M. P., Gandomzadeh A., Lenti L., Semblat J-F., Bonilla F., and S. Martino [2010], "Nonlinear site effects: interest of one directional –tree component (1C-3D) formulation", *Proc. 5th International Conference on Recent Advances in Geotechnical Earthquake Engineering and Soil Dynamics*, may 24-29 in 2010, paper n° 3.12b, ISBN 1-887009-15-9, S. Diego (California).

Sun, J.I., Golesorkhi, R., and H.B. Seed [1988], "Dynamic moduli and damping ratios for cohesive soils", Report No. EERC-88/15, Earthquake Engineering Research Centre, University of California, Berkeley.

Tertulliani, A., Tosi. P. and V. De Rubeis [1996], "Local seismicity in Rome (Italy): recent results from macroseismic evidences", Annali di Geofisica, 39(6), 1235-1240.

Towhata, I., and K. Ishihara [1985]. "Modeling Soil Behavior Under Principal Axes Rotation" *Fifth International Conference on Numerical Methods in Geomechanics*, Nagoya, 523-530.

Virieux J. [1986], "P-SV wave propagation in heterogeneous media: velocity stress finite-difference method", Geophysics, 51, 4, 889-901.